\documentclass[12pt]{article}
\usepackage{epsfig}
\usepackage{psfrag}
\usepackage{latexsym}
\def\beq{\begin{equation}}
\def\eeq{\end{equation}}
\def\bea{\begin{eqnarray}}
\def\eea{\end{eqnarray}}
\def\bet{\begin{tabular}}
\def\eet{\end{tabular}}
\def\bes{\begin{subequations}\bea}
\def\ees{\eea\end{subequations}}

\def\Cc{\cos\chi}
\def\Sc{\sin\chi}
\def\CCc{\cos^2\chi}
\def\SSc{\sin^2\chi}
\def\Cp{\cos\varphi}
\def\Sp{\sin\varphi}
\def\CCp{\cos^2\varphi}
\def\SSp{\sin^2\varphi}

\def\dza{\delta z_1}
\def\dzb{\delta z_2}
\def\dzc{\delta z_3}
\def\dze{\delta z_4}
\def\dzf{\delta z_5}
\def\be{\begin{equation}}
\def\ee{\end{equation}}
\def\bc{\begin{center}}
\def\ec{\end{center}}
\def\bea{\begin{eqnarray}}
\def\eea{\end{eqnarray}}
\def\dd{\displaystyle}
\def\nn{\nonumber}

\catcode`@=11
\def\marginnote#1{}
\newcount\hour
\newcount\minute
\newtoks\amorpm
\hour=\time\divide\hour by60
\minute=\time{\multiply\hour by60 \global\advance\minute by-\hour}
\edef\standardtime{{\ifnum\hour<12 \global\amorpm={am}%
        \else\global\amorpm={pm}\advance\hour by-12 \fi
        \ifnum\hour=0 \hour=12 \fi
        \number\hour:\ifnum\minute<10 0\fi\number\minute\the\amorpm}}
\edef\militarytime{\number\hour:\ifnum\minute<10 0\fi\number\minute}
\def\draftlabel#1{{\@bsphack\if@filesw {\let\thepage\relax
   \xdef\@gtempa{\write\@auxout{\string
      \newlabel{#1}{{\@currentlabel}{\thepage}}}}}\@gtempa
   \if@nobreak \ifvmode\nobreak\fi\fi\fi\@esphack}
        \gdef\@eqnlabel{#1}}
\def\@eqnlabel{}
\def\@vacuum{}
\def\draftmarginnote#1{\marginpar{\raggedright\scriptsize\tt#1}}
\def\draft{\oddsidemargin 0.0truein
        \def\@oddfoot{\sl preliminary draft \hfil
        \rm\thepage\hfil\sl\today\quad\militarytime}
        \let\@evenfoot\@oddfoot \overfullrule 3pt
        \let\label=\draftlabel
        \let\marginnote=\draftmarginnote
   \def\@eqnnum{(\theequation)\rlap{\kern\marginparsep\tt\@eqnlabel}%
\global\let\@eqnlabel\@vacuum}  }
\catcode`@=12
\begin{document}
\begin{titlepage}
\vspace*{-1cm}
\phantom{hep-ph/***} 
\hfill{DFPD-05/TH/22}

\hfill{CERN-PH-TH/2005-226}

\vskip 2.5cm
\begin{center}
{\Large\bf Tri-Bimaximal Neutrino Mixing, $A_4$ and the Modular Symmetry}
\end{center}
\vskip 0.2  cm
\vskip 0.5  cm
\begin{center}
{\large Guido Altarelli}~\footnote{e-mail address: guido.altarelli@cern.ch}
\\
\vskip .1cm
CERN, Department of Physics, Theory Division
\\ 
CH-1211 Geneva 23, Switzerland
\\
\vskip .1cm
and
\\
Dipartimento di Fisica `E.~Amaldi', Universit\`a di Roma Tre
\\ 
INFN, Sezione di Roma Tre, I-00146 Rome, Italy
\\
\vskip .2cm
{\large Ferruccio Feruglio}~\footnote{e-mail address: feruglio@pd.infn.it}
\\
\vskip .1cm
Dipartimento di Fisica `G.~Galilei', Universit\`a di Padova 
\\ 
INFN, Sezione di Padova, Via Marzolo~8, I-35131 Padua, Italy
\\
\end{center}
\vskip 0.7cm
\begin{abstract}
\noindent
We formulate and discuss a 4-dimensional SUSY version of an $A_4$ model for tri-bimaximal neutrino mixing which is completely natural. We also study the next-to-the-leading corrections and show that they are small, once the ratios of $A_4$ breaking VEVs to the cutoff are fixed in a specified interval. We also point out an interesting way of presenting the $A_4$ group starting from the modular group. In this approach, which could be interesting in itself as an indication on a possible origin of $A_4$, the lagrangian basis where the symmetry is formulated coincides with the basis where the charged leptons are diagonal. If the same  classification structure in $A_4$ is extended from leptons to quarks,  the CKM matrix coincides with the unit matrix in leading order and a study of non leading corrections shows that the departures from unity of the CKM matrix are far too small to accomodate the observed mixing angles. 

\end{abstract}
\end{titlepage}
\setcounter{footnote}{0}
\vskip2truecm
%
\section{Introduction}
It is an experimental fact \cite{data} that within measurement errors the observed form of the neutrino mixing matrix is compatible with the so called tri-bimaximal form, discussed by Harrison, Perkins and Scott (HPS) \cite{hps}, which, apart from phase redefinitions is given by: 
\be
U=\left(
\begin{array}{ccc}
\sqrt{2/3}& 1/\sqrt{3}& 0\\
-1/\sqrt{6}& 1/\sqrt{3}& -1/\sqrt{2}\\
-1/\sqrt{6}& 1/\sqrt{3}& +1/\sqrt{2}
\end{array}
\right)~~~.
\ee
It is an interesting challenge to formulate dynamical principles that can lead to this specific mixing pattern in a completely natural way as a first approximation, with small corrections determined by higher order terms in a well defined expansion. In a series of papers \cite{ma1,ma1.5,ma2} it has been pointed out that a broken flavour symmetry based on the discrete group $A_4$ appears to be particularly fit for this purpose. Other solutions based on continuous flavour groups like SU(3) or SO(3) have also been recently presented \cite{continuous,others}, but the $A_4$ models have a very simple (for example, in terms of field content) and attractive structure. 
In a recent paper \cite{OurTriBi} we have constructed an explicit $A_4$ model where the problem as stated above is solved. A crucial feature of all HPS models is the mechanism used to guarantee the necessary VEV alignment of the flavon field $\varphi_T$ which determines the charged lepton mass matrix with respect to the
direction in flavour space chosen by the flavon $\varphi_S$ that gives the neutrino mass matrix. In ref. \cite{OurTriBi} we adopted an extra dimensional framework, with $\varphi_T$ and $\varphi_S$ on different branes so that the minimization of the respective potentials is kept to a large extent independent. The advantage of this approach is that the HPS mixing is reproduced quite naturally in a comparatively simple way. A moderately hierarchical neutrino mass spectrum is obtained. The correction terms from 
all possible higher dimensionality operators allowed by the symmetries of the model are shown to be small for a wide range of values for the cut-off 
scales. Finally, the observed hierarchy 
of charged lepton masses can be reproduced without fine tuning by enlarging the flavour symmetry with an extra U(1). 

In the present article we address a number of questions which are left open. First, we  give an alternative formulation of the $A_4$ model in 4 dimensions with supersymmetry (SUSY) which shows that the connection of $A_4$ with the HPS matrix is robust, in the sense that it can be obtained in different ways and does not necessarily require extra dimensions. The two versions differ in the set of additional fields and in the pattern of non leading corrections, so that experimental tests are in principle possible. The advantage of SUSY is to considerably simplify the problem of obtaining the right vacuum alignment from the minimization of the relevant potential. We present a detailed discussion of the pattern of non leading corrections in the new version of the model. We also address the important problem of trying to understand the dynamical origin of $A_4$. To this end we reformulate the 12 elements of $A_4$ as products of two matrices $S$ and $T$ with $S^2=(ST)^3=T^3=1$. In this formulation the lagrangian basis directly coincides with that where the charged leptons are diagonal. The main virtue of this formulation is that $A_4$ is seen as a subgroup of the modular group of trasformations which often plays a role in the formalism of string theories, for example in the context of duality trasformations \cite{gpr}.  
We then discuss the extension to quarks. We show that a direct extrapolation to quarks of the classification scheme adopted in $A_4$ for leptons immediately leads in lowest approximation to a diagonal CKM mixing matrix. This fact had been already observed in a similar $A_4$ context in refs. \cite{ma1,ma1.5}. We study the higher order corrections in our specific framework and show that  a direct extension to the quark sector of the $A_4$ classification leads, apart from negligible terms, to the same contributions for up and down mass matrices, so that the CKM matrix remains diagonal. Thus new sources of $A_4$ breaking are needed in the quark sector.

%
%
\section{$A_4$ revisited}

$A_4$ can also be defined as the group generated by the two elements
$S$ and $T$ obeying the relations \cite{presentation} (a "presentation" of the group):
\be
S^2=(ST)^3=T^3=1~~~.
\label{$A_4$}
\ee
It is immediate to see that one-dimensional unitary representations are
given by:
\be
\begin{array}{lll}
1&S=1&T=1\\
1'&S=1&T=e^{\dd i 4 \pi/3}\equiv\omega^2\\
1''&S=1&T=e^{\dd i 2\pi/3}\equiv\omega\\
\label{s$A_4$}
\end{array}
\ee
It is simple to check that a two-dimensional unitary representation does not exist (only $\det(T^3)=-1$ is in this case compatible with $S^2=(ST)^3=1$).
The three-dimensional unitary representation, in a basis
where the element $T$ is diagonal, is given by:
\be
T=\left(
\begin{array}{ccc}
1&0&0\\
0&\omega^2&0\\
0&0&\omega
\end{array}
\right),~~~~~~~~~~~~~~~~
S=\frac{1}{3}
\left(
\begin{array}{ccc}
-1&2&2\cr
2&-1&2\cr
2&2&-1
\end{array}
\right)~~~.
\label{ST}
\ee
The elements of $A_4$  can be represented by the $3\times 3$ matrices $M_i$ ($i=1,2,...12$) given by:
$1$, $S$, $T$, $ST$, $TS$, $T^2$, $ST^2$, $STS$, $TST$, $T^2S$, $TST^2$, $T^2ST$. 
Indeed if one performs the unitary transformation $M_i \rightarrow M'_i = V^\dagger M_i V$ with 
\be
V=\frac{1}{\sqrt{3}}\left(
\begin{array}{ccc}
1& 1& 1\\
1& \omega^2& \omega\\
1& \omega& \omega^2
\end{array}
\right)~~~,
\label{change}
\ee
one obtains the 12 matrices listed in eqs. (5-8) of ref.\cite{ma1} which span the three-dimensional representation of $A_4$.

Starting from the explicit expressions of $S$ and $T$ we can build
the multiplication rules for triplet representations. Consider the two triplets:
\be
a=(a_1,a_2,a_3)~~~,~~~~~~~~~~b=(b_1,b_2,b_3)~~~.
\ee
The combination $x a_1 b_1+y a_2 b_3+z a_3 b_2$ is invariant under $T$. If we also enforce invariance under
$S$ it is easy to see that we get the invariant singlet
\be
1\equiv(ab)=(a_1 b_1+a_2 b_3+a_3 b_2)~~~.
\label{dec1}
\ee
Similarly, the singlets $1'$ and $1''$ are obtained from the combinations
$x' a_3 b_3+y' a_1 b_2+z' a_2 b_1$ and $x'' a_2 b_2+y'' a_3 b_1+z'' a_1 b_3$, by imposing invariance under $S$:
\be
\begin{array}{llll}
1'&\equiv(ab)'&=&(a_3 b_3+a_1 b_2+a_2 b_1)\\
1''&\equiv(ab)''&=&(a_2 b_2+a_1 b_3+a_3 b_1)
\label{dec2}
\end{array}
\ee
With a little algebra it is possible to see that the remaining 6 independent combinations
fill two triplets, a symmetric one and an antisymmetric one:
\bea
3\equiv(ab)_S&=&\frac{1}{3}(2 a_1 b_1-a_2 b_3-a_3 b_2,2 a_3 b_3-a_1 b_2-a_2 b_1,2 a_2 b_2-a_1 b_3-a_3 b_1)\cr
3'\equiv(ab)_A&=&\frac{1}{2}(a_2 b_3-a_3 b_2,a_1 b_2-a_2 b_1,a_1 b_3-a_3 b_1)
\label{dec3}
\eea
Moreover, if $c$, $c'$ and $c''$ are singlets of the type $1$, $1'$ and $1''$, and 
$a=(a_1,a_2,a_3)$ is a triplet, then the products $ac$, $ac'$ and $ac''$ are triplets
explicitly given by $(a_1 c,a_2 c, a_3 c)$, $(a_3 c',a_1 c', a_2 c')$ and $(a_2 c'',a_3 c'', a_1 c'')$,
respectively. 

The group $A_4$ has two obvious subgroups: $G_S$, which is a reflection subgroup
generated by $S$ and $G_T$, which is the group generated by $T$, isomorphic to $Z_3$.
If the flavour symmetry associated to $A_4$ is broken by the VEV of a triplet
$\varphi=(\varphi_1,\varphi_2,\varphi_3)$ of scalar fields,
there are two interesting breaking pattern. The VEV
\be
\langle\varphi\rangle=(1,1,1)
\label{unotre}
\ee
breaks $A_4$ down to $G_S$, while
\be
\langle\varphi\rangle=(1,0,0)
\label{unozero}
\ee
breaks $A_4$ down to $G_T$. As we will see, $G_S$ and $G_T$ are the relevant low-energy symmetries
of the neutrino and the charged-lepton sectors, respectively. Note that the vectors (\ref{unotre}) and 
(\ref{unozero}) are interchanged under the transformation in eq. (\ref{change}).

%
\section{Basic Structure of the Model}

We discuss here general properties of the model based on the $A_4$ realization discussed above which are independent of the particular mechanism adopted to guarantee the required VEV alignment which will be specified in the next section. Following ref. \cite{OurTriBi} we assigns leptons to the four inequivalent
representations of the group $A_4$: left-handed lepton doublets $l$ transform
as a triplet $3$, while the right-handed charged leptons $e^c$,
$\mu^c$ and $\tau^c$ transform as $1$, $1''$ and $1'$, respectively. 
The flavour symmetry is broken by two triplets
$\varphi_S$ and $\varphi_T$ and by a singlet $\xi$. Actually we may need more singlets and indeed in the next section we will introduce two of them. But we can always choose a basis
in the space of these singlets such that $\xi$ denotes the field
with a non-vanishing VEV, whereas all the other ones have a
zero VEV and do not contribute to the neutrino mass matrices. 
So in this section we only keep the terms with $\xi$ for simplicity.
All these fields are gauge singlets.
Two Higgs doublets $h_{u,d}$, invariant under $A_4$, are
also introduced. 
We assume that some mechanism produces and maintains the hierarchy
$\langle h_{u,d}\rangle=v_{u,d}\ll \Lambda$ where $\Lambda$ is the 
cut-off scale of the theory (for example, by adopting a supersymmetric version of the model, as in the next section).
The Yukawa interactions in the lepton sector read:
\be
{\cal L}_{l}=y_e e^c (\varphi_T l)+y_\mu \mu^c (\varphi_T l)'+
y_\tau \tau^c (\varphi_T l)''+ x_a\xi (ll)+x_b (\varphi_S ll)+h.c.+...
\label{wl}
\ee 
To keep our formulae compact, we use a two-component notation
for the fermion fields and omit to write the Higgs fields
$h_{u,d}$ and the cut-off scale $\Lambda$. For instance 
$y_e e^c (\varphi_T l)$ stands for $y_e e^c (\varphi_T l) h_d/\Lambda$,
$x_a\xi (ll)$ stands for $x_a\xi (l h_u l h_u)/\Lambda^2$ and so on.
The Lagrangian  ${\cal L}_l$ contains the lowest order operators
in an expansion in powers of $1/\Lambda$. Dots stand for higher
dimensional operators that will be discussed later on. 
Some terms allowed by the flavour symmetry, such as the terms 
obtained by the exchange $\varphi_T \leftrightarrow \varphi_S$, 
or the term $(ll)$ are missing in ${\cal L}_l$. 
Their absence is crucial and will be
motivated later on.
As we will demonstrate, the fields $\varphi_T$,
$\varphi_S$ and $\xi$ develop a VEV along the directions:
\bea
\langle \varphi_T \rangle&=&(v_T,0,0)\nn\\ 
\langle \varphi_S\rangle&=&(v_S,v_S,v_S)\nn\\
\langle \xi \rangle&=&u~~~. 
\label{align}
\eea 
At the leading order in $1/\Lambda$ and after the breaking of the 
flavour and electroweak symmetries, the mass terms from the Lagrangian 
(\ref{wl}) are 
\bea
{\cal L}_l&=& v_d\frac{v_T}{\Lambda}\left(y_e e^c e+y_\mu \mu^c \mu+
y_\tau \tau^c \tau\right)\nn\\
&+&
x_av_u^2\frac{u}{\Lambda^2} (\nu_e\nu_e+2\nu_\mu \nu_\tau)\nn\\
&+&x_bv_u^2\frac{2v_S}{3\Lambda^2} 
(\nu_e\nu_e+\nu_\mu\nu_\mu+\nu_\tau\nu_\tau-\nu_e\nu_\mu-\nu_\mu\nu_\tau-\nu_\tau\nu_e)+h.c.+...
\label{wlmasses}
\eea
Here we have made use of eqs. (\ref{dec1},\ref{dec2},\ref{dec3}).
In the charged lepton sector the flavour symmetry $A_4$ is broken by $\langle \varphi_T \rangle$ down to
$G_T$. Actually the above mass terms for charged leptons are the most general allowed by the 
symmetry $G_T$. At leading order in $1/\Lambda$, charged lepton masses are diagonal simply because
there is a low-energy $G_T$ symmetry. In the neutrino sector $A_4$ is broken down to $G_S$,
though neutrino masses in this model are not the most general ones allowed by $G_S$.
From eq. (\ref{wlmasses}), at the leading order of the $1/\Lambda$ expansion,
we read the mass matrices $m_l$ and $m_\nu$ for charged leptons and 
neutrinos:
\be
m_l=v_d\frac{v_T}{\Lambda}\left(
\begin{array}{ccc}
y_e& 0& 0\\
0& y_\mu & 0 \\
0& 0& y_\tau 
\end{array}
\right)~~~,
\label{mch}
\ee
\be
m_\nu=\frac{v_u^2}{\Lambda}\left(
\begin{array}{ccc}
a+2 b/3& -b/3& -b/3\\
-b/3& 2b/3& a-b/3\\
-b/3& a-b/3& 2 b/3
\end{array}
\right)~~~,
\label{mnu}
\ee
where 
\be
a\equiv 2 x_a\frac{u}{\Lambda}~~~,~~~~~~~b\equiv 2 x_b\frac{v_S}{\Lambda}~~~.
\label{ad}
\ee
Charged fermion masses are given by:
\be
m_e= y_e v_d \frac{v_T}{\Lambda}~~~,~~~~~~~
m_\mu=y_\mu v_d \frac{v_T}{\Lambda}~~~,~~~~~~~
m_\tau= y_\tau v_d \frac{v_T}{\Lambda}~~~.
\label{chmasses}
\ee
We can easily obtain a natural hierarchy among $m_e$, $m_\mu$ and
$m_\tau$ by introducing an additional U(1)$_F$ flavour symmetry under
which only the right-handed lepton sector is charged.
We write the F-charge values in this model  as $0$, $q$ and $2q$ for $\tau^c$, $\mu^c$ and
$e^c$, respectively. By assuming that a flavon $\theta$, carrying
a negative unit of F, acquires a VEV 
$\langle \theta \rangle/\Lambda\equiv\lambda<1$, the Yukawa couplings
become field dependent quantities $y_{e,\mu,\tau}=y_{e,\mu,\tau}(\theta)$
and we have
\be
y_\tau\approx O(1)~~~,~~~~~~~y_\mu\approx O(\lambda^q)~~~,
~~~~~~~y_e\approx O(\lambda^{2q})~~~\label{q}.
\ee
The neutrino mass matrix is diagonalized by the transformation:
\be
U^T m_\nu U =\frac{v_u^2}{\Lambda}{\tt diag}(m_1=a+b,m_2=a,m_3=-a+b)~~~,
\ee
with
\be
U=\left(
\begin{array}{ccc}
\sqrt{2/3}& 1/\sqrt{3}& 0\\
-1/\sqrt{6}& 1/\sqrt{3}& -1/\sqrt{2}\\
-1/\sqrt{6}& 1/\sqrt{3}& +1/\sqrt{2}
\end{array}
\right)~~~.
\label{HPSmatrix}
\ee
Thus the HPS mixing matrix is obtained in the leading approximation. The constraints on the parameters and on the scales of the model in order to obtain a realistic neutrino mass spectrum are exactly as in our previous paper \cite{OurTriBi}. In fact the formulation of $A_4$ adopted in this paper is such that the lagrangian basis where the symmetry is specified coincides with the basis where the charged leptons are diagonal. This makes the role of $A_4$ in producing the HPS mixing more transparent but, at the leading level, the old and the new versions are related by a unitary change of basis. Thus the present version of   $A_4$ model is completely natural, as much as our previous version with an extra dimension, both needing only a moderate amount of fine tuning to reproduce the small value of $r=\Delta m_{sun}^2/\Delta m_{atm}^2 $. The difference between the extra dimensional version and the present SUSY version is however relevant at the level of subleading terms.
We recall the expected range for the parameters in the symmetry breaking sector. By assuming that all the VEVs breaking
$A_4$ have approximately the same value, in ref. \cite{OurTriBi} we found that
\be
0.004< \frac{v_S}{\Lambda}\approx \frac{v_T}{\Lambda}\approx \frac{u}{\Lambda}<1
\label{range}
\ee
and that the cutoff $\Lambda$ should be limited between $10^{13}$ GeV and $2\times 10^{15}$ GeV.
In particular, the lower bound in (\ref{range}) comes from requiring that the tau Yukawa coupling is within a perturbative regime. In the supersymmetric version of the model discussed in the next section, the tau Yukawa coupling
is given by $y_\tau=(m_\tau \Lambda)/(v \cos\beta v_T)$ where $v\approx 174$ GeV and $\tan\beta=v_u/v_d$.
By asking $y_\tau<4 \pi$ we find $v_T/\Lambda>0.0022(0.024)$ when $\tan\beta=2.5(30)$. In what follows
we take as lower limit $v_T/\Lambda>0.0022$.

In conclusion, if one can  construct a natural mechanism to guarantee the necessary alignment of the VEVs $\varphi_S$ and $\varphi_T$, one obtains a first approximation where the neutrino mixing is of the HPS form.
In the next section we will present a supersymmetric version of the model in four dimensions and later we will discuss the non leading corrections in this context.

%
%
\section{Vacuum Alignment}
Here we discuss a supersymmetric solution to the vacuum alignment problem.
In a SUSY context, the right-hand side of eq. (\ref{wl})
should be interpreted as the superpotential $w_l$ of the theory,
in the lepton sector:
\be
w_l=y_e e^c (\varphi_T l)+y_\mu \mu^c (\varphi_T l)'+
y_\tau \tau^c (\varphi_T l)''+ (x_a\xi+\tilde{x}_a\tilde{\xi}) (ll)+x_b (\varphi_S ll)+h.c.+...
\label{wlplus}
\ee
where dots stand for higher dimensional operators that will be discussed
in the next section and where we have also added an additional 
$A_4$-invariant singlet
$\tilde{\xi}$. Such a singlet does not modify the structure of the mass matrices
discussed previously, but plays an important role in the vacuum 
alignment mechanism.
A key observation is that the superpotential $w_l$ 
is invariant not only with respect to the gauge symmetry 
SU(2)$\times$ U(1) and the flavour symmetry U(1)$_F\times A_4$,
but also under a discrete $Z_3$ symmetry and a continuous U(1)$_R$ 
symmetry under which the fields 
transform as shown in the following table.
\\[0.2cm]
\begin{center}
\begin{tabular}{|c||c|c|c|c||c|c|c|c|c||c|c|c|}
\hline
{\tt Field}& l & $e^c$ & $\mu^c$ & $\tau^c$ & $h_{u,d}$ & 
$\varphi_T$ & $\varphi_S$ & $\xi$ & $\tilde{\xi}$ & $\varphi_0^T$ & $\varphi_0^S$ & $\xi_0$\\
\hline
$A_4$ & $3$ & $1$ & $1'$ & $1''$ & $1$ & 
$3$ & $3$ & $1$ & $1$ & $3$ & $3$ & $1$\\
\hline
$Z_3$ & $\omega$ & $\omega^2$ & $\omega^2$ & $\omega^2$ & $1$ &
$1$ & $\omega$ & $\omega$ & $\omega$ & $1$ & $\omega$ & $\omega$\\
\hline
$U(1)_R$ & $1$ & $1$ & $1$ & $1$ & $0$ & 
$0$ & $0$ & $0$ & $0$ & $2$ & $2$ & $2$\\
\hline
\end{tabular}
\end{center}
\vspace{0.2cm}
We see that the $Z_3$ symmetry explains the absence of the term $(ll)$
in $w_l$: such a term transforms as $\omega^2$ under $Z_3$ and
need to be compensated by the field $\xi$ in our construction.
At the same time $Z_3$ does not allow the interchange between
$\varphi_T$ and $\varphi_S$, which transform differently under $Z_3$. 
The singlets $\xi$ and $\tilde{\xi}$ have the same transformation properties
under all symmetries and, as we shall see, in a finite range of parameters,
the VEV of $\tilde{\xi}$ vanishes and does not contribute to neutrino masses. 
Charged leptons and neutrinos acquire
masses from two independent sets of fields. 
If the two sets of fields develop VEVs according to the 
alignment described in eq. (\ref{align}), then the desired
mass matrices follow.

Finally, there is a continuous $U(1)_R$
symmetry that contains the usual $R$-parity as a subgroup.
Suitably extended to the quark sector, this symmetry forbids
the unwanted dimension two and three terms in the superpotential
that violate baryon and lepton number at the renormalizable level. 
The $U(1)_R$ symmetry allows us to classify
fields into three sectors. There are ``matter fields'' such as the 
leptons $l$, $e^c$, $\mu^c$ and $\tau^c$, which occur in the 
superpotential through bilinear combinations. There is a 
``symmetry breaking sector'' including the higgs doublets
$h_{u,d}$ and the flavons $\varphi_T$, $\varphi_S$, $(\xi,\tilde{\xi})$.
As we will see these fields acquire non-vanishing VEVs and break the symmetries of the model.
Finally, there are ``driving fields'' such as $\varphi_0^T$, $\varphi_0^S$
and $\xi_0$ that allows to build a 
non-trivial scalar potential in the symmetry breaking sector. 
Since driving fields have R-charge equal to two, the superpotential
is linear in these fields.

The full superpotential of the model is
\be
w=w_l+w_d
\ee 
where, at leading order in a $1/\Lambda$ expansion, $w_l$ is given
by the right-hand side of eq. (\ref{wl}) and the ``driving'' term 
$w_d$ reads:
\bea
w_d&=&M (\varphi_0^T \varphi_T)+ g (\varphi_0^T \varphi_T\varphi_T)\nn\\
&+&g_1 (\varphi_0^S \varphi_S\varphi_S)+
g_2 \tilde{\xi} (\varphi_0^S \varphi_S)+
g_3 \xi_0 (\varphi_S\varphi_S)+
g_4 \xi_0 \xi^2+
g_5 \xi_0 \xi \tilde{\xi}+
g_6 \xi_0 \tilde{\xi}^2~~~.
\label{wd}
\eea
At this level there is no fundamental distinction between the singlets
$\xi$ and $\tilde{\xi}$. Thus we are free to define $\tilde{\xi}$ as the combination
that couples to $(\varphi_0^S \varphi_S)$ in the superpotential $w_d$.
We notice that at the leading order there are no terms involving
the Higgs fields $h_{u,d}$. We assume that the electroweak symmetry
is broken by some mechanism, such as radiative effects when SUSY is broken. It is interesting that at the leading order
the electroweak scale does not mix with the potentially large scales
$u$, $v$ and $v'$. The scalar potential is given by:
\be
V=\sum_i\left\vert\frac{\partial w}{\partial \phi_i}\right\vert^2
+m_i^2 \vert \phi_i\vert^2+...
\ee
where $\phi_i$ denote collectively all the scalar fields of the 
theory, $m_i^2$ are soft masses and dots stand for D-terms for the 
fields charged under the gauge group and possible additional
soft breaking terms. Since $m_i$ are expected to be much smaller
than the mass scales involved in $w_d$, it makes sense to
minimize $V$ in the supersymmetric limit and to account for soft 
breaking effects subsequently. From the driving sector we have:
\bea
\frac{\partial w}{\partial \varphi^T_{01}}&=&M{\varphi_T}_1
+\frac{2 g}{3}({\varphi_T}_1^2-{\varphi_T}_2{\varphi_T}_3)=0\nn\\
\frac{\partial w}{\partial \varphi^T_{02}}&=&M{\varphi_T}_3
+\frac{2 g}{3}({\varphi_T}_2^2-{\varphi_T}_1{\varphi_T}_3)=0\nn\\
\frac{\partial w}{\partial \varphi^T_{03}}&=&M{\varphi_T}_2
+\frac{2 g}{3}({\varphi_T}_3^2-{\varphi_T}_1{\varphi_T}_2)=0\nn\\
\frac{\partial w}{\partial \varphi^S_{01}}&=&g_2\tilde{\xi} {\varphi_S}_1+
\frac{2g_1}{3}({\varphi_S}_1^2-{\varphi_S}_2{\varphi_S}_3)=0\nn\\
\frac{\partial w}{\partial \varphi^S_{02}}&=&g_2\tilde{\xi} {\varphi_S}_3+
\frac{2g_1}{3}({\varphi_S}_2^2-{\varphi_S}_1{\varphi_S}_3)=0\nn\\
\frac{\partial w}{\partial \varphi^S_{03}}&=&g_2\tilde{\xi} {\varphi_S}_2+
\frac{2g_1}{3}({\varphi_S}_3^2-{\varphi_S}_1{\varphi_S}_2)=0\nn\\
\frac{\partial w}{\partial \xi_0}&=&
g_4 \xi^2+g_5 \xi \tilde{\xi}+g_6\tilde{\xi}^2
+g_3({\varphi_S}_1^2+2{\varphi_S}_2{\varphi_S}_3)=0
\eea
A solution to the first three equations is:
\be
\varphi_T=(v_T,0,0)~~~,~~~~~~~v_T=-\frac{3M}{2g}~~~.
\label{solT}
\ee
This VEV breaks $A_4$ down to $G_T$
\footnote{More precisely, since the solutions lie in an orbit 
of the group $A_4$, the non trivial solutions are
(\ref{solT}) and those generated by acting on (\ref{solT})
by the elements of $A_4$: $\varphi_T=(M/2g)(1,-2,-2)$,
$\varphi_T=(M/2g)(1,-2\omega^2,-2\omega)$ and 
$\varphi_T=(M/2g)(1,-2\omega,-2\omega^2)$.
Each of these vacua leaves unbroken a $Z_3$ subgroup of $A_4$.
It is not restrictive to choose the vacuum $\varphi_T=-(3M/2g)(1,0,0)$.
The trivial solution $\varphi_T=(0,0,0)$
can be eliminated by choosing $m_{\varphi_T}^2<0$.}.
The need of an additional singlet can be understood by looking at
the remaining equations. Indeed, if a unique singlet were present,
which can be realized by setting $\xi=0$, then the only solution to
these equations would be that with all vanishing VEVs for $\varphi_S$
and $\tilde{\xi}$. The additional singlet is therefore essential to
recover a non-trivial solution
\footnote
{We are indebted to G.G. Ross for pointing out this possibility to us.}.
In particular, if we choose $m^2_{\tilde{\xi}}>0$, thus enforcing 
$\langle \tilde{\xi}\rangle=0$, in a finite portion of the parameter space
we find the solution
\bea
\tilde{\xi}&=&0\nn\\
\xi&=&u\nn\\
\varphi_S&=&(v_S,v_S,v_S)~~~,~~~~~~~~~v_S^2=-\frac{g_4}{3 g_3} u^2~~~
\label{solS}
\eea
with $u$ undetermined 
\footnote{Also in this case we find other degenerate solutions,
obtained by acting on (\ref{solS}) with the elements of $A_4$:
$\varphi_S=v_S(1,\omega,\omega^2)$ and $\varphi_S=v_S(1,\omega^2,\omega)$.
Any of these solutions produces the same neutrino mass matrix.
For instance $\varphi_S=v_S(1,\omega,\omega^2)$ is equivalent to 
$\varphi_S=v_S(1,1,1)$, the two being related by the local
field transformations $\nu_e\to \nu_e,\nu_\mu\to \omega^2 \nu_\mu, 
\nu_\tau\to \omega \nu_\tau$.}.
By choosing $m^2_{\varphi_0},m^2_{\varphi'_0},
m^2_{\xi_0}>0$, the driving fields $\varphi_0$, $\varphi'_0$
and $\xi_0$ vanish at the minimum. Moreover, if $m^2_{\varphi_S},
m^2_\xi<0$, then $u$ slides to a large scale, which we assume to be eventually
stabilized by one-loop radiative corrections.

As for the U(1)$_F$ field $\theta$ it is easy to see that, in the unbroken SUSY limit, its VEV remains undetermined as a consequence of the vanishing of the charged lepton field VEVs. When SUSY is broken
the $m_\theta^2 |\theta|^2$ term in the potential would drive the $\theta$ VEV to zero (or to $\infty$ if $m_\theta^2 < 0$). However, in the presence of a renormalizable coupling of $\theta$ to additional field(s), like, for example, $g_\sigma \theta \sigma^2$, one-loop radiative corrections typically bring back the $\theta$ VEV near the cutoff. We implicitly assume that such field(s) $\sigma$,...., which are completely neutral except for the appropriate U(1)$_F$ and U(1)$_R$ charges, are included in our model. Thus the value of the ratio $\langle \theta/ \rangle \Lambda$ can be taken as a free parameter that, together with the charge value $q$, fixes the charged lepton mass ratios as given in eq. (\ref{q}).
\vspace{0.5cm}
%
\section{Higher-order corrections}
The results of the previous section hold to first approximation.
Higher-dimensional operators, suppressed by additional powers of 
the cut-off $\Lambda$, can be added to the leading terms in the lagrangian.
Here we will classify these non leading terms and analyze their
physical effects.  In particular we will show that these
corrections are completely under control in our model and that
they can be made negligibly small without any fine-tuning.
We can classify higher-order operators
into three groups.
\subsection{Corrections to $m_l$}
The leading operators giving rise to $m_l$ are of order
$1/\Lambda$ (see eqs. (\ref{wlmasses},\ref{mch})). At order $(1/\Lambda)^2$ there are no new structures
contributing to $m_l$. Indeed the only invariant operator:
\be
\frac{1}{\Lambda^2} (f^c l\varphi_T\varphi_T) h_d~~~,
~~~~~~(f^c=e^c,\mu^c,\tau^c)
\ee
replicates the leading-order pattern, as can be seen from the fact that
the symmetric triplet
\be
(\varphi_T\varphi_T)_S=\frac{2}{3}({\varphi_T}_1^2-{\varphi_T}_2{\varphi_T}_3,
{\varphi_T}_3^2-{\varphi_T}_1{\varphi_T}_2,{\varphi_T}_2^2-{\varphi_T}_3
{\varphi_T}_1)
\ee
has a VEV in the same direction as $\varphi_T$. Thus, in the charged lepton
sector, the first corrections arise at relative order $1/\Lambda^2$.
\subsection{Corrections to $m_\nu$}
The leading operators contributing to $m_\nu$ are of order $1/\Lambda^2$ (see eqs. (\ref{wlmasses},\ref{mnu})).
At the next order we have three operators, whose contribution
to $m_\nu$, after symmetry breaking, 
cannot be absorbed by a redefinition of the parameters
$x_{a,b}$:
\be
\begin{array}{l}
\dd\frac{x_c}{\Lambda^3}(\varphi_T\varphi_S)' (ll)'' h_uh_u\\
\\
\dd\frac{x_d}{\Lambda^3}(\varphi_T\varphi_S)'' (ll)' h_uh_u\\
\\
\dd\frac{x_e}{\Lambda^3}\xi (\varphi_T ll) h_uh_u
\\
\end{array}~~~.
\label{hbc}
\ee
The corrections from these operators will be taken into account in Sect. 5.4.

\subsection{Corrections to the vacuum alignment}
In the appendix B, the operators of higher dimension contributing to the superpotential $w_d$ introduced in eq. (\ref{wd}) are listed and the procedure of minimization is repeated. The leading corrections to the VEVs are of relative order $1/\Lambda$ and affect all the flavon fields.  The correction to 
the VEV of  $\varphi_T$, apart from a shift of $v_T$, is proportional to the VEV of $\varphi_S$. In turn,
the VEV of $\varphi_S$ is shifted in a generic direction, the VEV of $\tilde{\xi}$, which was vanishing at leading order, acquires a small component 
and that of $\xi$ remains undetermined:
\be
\begin{array}{ccl}
\langle \varphi_T\rangle& \to& (v_T'+\delta v_T, \delta v_T, \delta v_T)\\
\langle \varphi_S \rangle& \to& (v_S+\delta v_1,v_S+\delta v_2, v_S+\delta v_3)\\
\langle \xi \rangle&\to &u\\
\langle \tilde{\xi} \rangle& \to& \delta u'
\end{array}
\label{alisub}
\ee
where $u$ is undetermined, $v_T'-v_T$, $\delta v_T$, $\delta v_i$ and $\delta u'$ are suppressed with respect to $v_T$
and $v_S$ by a factor $1/\Lambda$.
\subsection{Modified masses and mixing angles}
The new vacuum in eq. (\ref{alisub}) modifies the leading order mass matrix $m_l$ in eq. (\ref{mch})
into $m_l'=m_l(v_T\to v_T')+\delta m_l$
\be
\delta m_l=v_d\frac{\delta v_T}{\Lambda}\left(
\begin{array}{ccc}
y_e& y_e& y_e\\
y_\mu& y_\mu & y_\mu \\
y_\tau& y_\tau& y_\tau 
\end{array}
\right)~~~.
\label{dmch}
\ee
The matrix ${m_l'}^\dagger m_l'$ is diagonalized by sending $l$ into $U_e l$
where, by neglecting terms of relative order $(y_e/y_\tau)^2$ and $(y_\mu/y_\tau)^2$, $U_e$
is given by:
\be
U_e=\left(
\begin{array}{ccc}
1&\dd\frac{\delta v_T}{v_T}& \dd\frac{\delta v_T}{v_T}\\
-\dd\frac{\delta v_T}{v_T}&1&\dd\frac{\delta v_T}{v_T}\\
-\dd\frac{\delta v_T}{v_T}&-\dd\frac{\delta v_T}{v_T}&1
\end{array}
\right)~~~.
\label{Ue}
\ee
Charged lepton masses are modified by an overall factor $(v_T'+\delta v_T)/v_T$.

The neutrino mass matrix is modified by both the new vacuum, eq. (\ref{alisub}), and by the new operators (\ref{hbc}). 
The non-vanishing VEV of $\tilde{\xi}$ can be absorbed into a redefinition of the parameter $a$ of
eq. (\ref{ad}). The remaining effects change $m_\nu$ of eq. (\ref{mnu})
into $m_\nu'=m_\nu+\delta m_\nu$
\be
\delta m_\nu=
\left(
\begin{array}{ccc}
2\dza/3 &-\dzc/3+\dzf &-\dzb/3+\dze\\
-\dzc/3+\dzf &2\dzb/3+\dze &-\dza/3\\
-\dzb/3+\dze&-\dza/3&2\dzc/3+\dzf
\end{array}
\right)\frac{v_u^2}{\Lambda}
\label{dmnu}
\ee
where 
\be
\dza\equiv 2 \left(\frac{x_b}{\Lambda} \delta v_1+ x_e\frac{u v_T}{\Lambda^2}\right)~~~,~~~~~~~
\dzb\equiv 2 \frac{x_b}{\Lambda} \delta v_2~~~,~~~~~~~
\dzc\equiv 2 \frac{x_b}{\Lambda} \delta v_3~~~,
\label{dz1}
\ee
\be
\dze\equiv 2 x_c\frac{v_S v_T}{\Lambda^2}~~~,~~~~~~~
\dzf\equiv 2 x_d\frac{v_S v_T}{\Lambda^2}~~~.
\label{dz2}
\ee
To first order in $\delta z_i$ neutrino masses are given by:
\be
\begin{array}{ccl}
m_1&=&\dd\frac{v_u^2}{\Lambda}\left[a+b+\dd\frac{1}{3}(\dza+\dzb+\dzc)
-\dd\frac{1}{2}(\dze+\dzf)\right]\\
m_2&=&\dd\frac{v_u^2}{\Lambda}\left[a+\dze+\dzf\right]\\
m_3&=&\dd\frac{v_u^2}{\Lambda}\left[-a+b+\dd\frac{1}{3}(\dza+\dzb+\dzc)
+\dd\frac{1}{2}(\dze+\dzf)\right]
\end{array}
\ee
By combining the first order corrections to neutrino and charged lepton masses
we find the modified parameters of the lepton mixing matrix (a bar on a letter indicates complex conjugation):
\be
\begin{array}{ccl}
\vert U_{e3}\vert&=&\dd\frac{1}{\sqrt{2}}
\left\vert\left[\dd\frac{1}{3(\bar{a} b+\bar{b} a-\vert b\vert^2)}
(\bar{a}(\dzb-\dzc)+(b-a)(\delta \bar{z}_2-\delta \bar{z}_3))\right.\right.\\
&-&\left.\left.\dd\frac{1}{2(\bar{a} b+\bar{b} a)}
((\bar{a}+\bar{b})(\dze-\dzf)-(a-b)(\delta\bar{z}_4-\delta\bar{z}_5))
\right]\right\vert
\end{array}
\label{ue3c}
\ee
\be
\tan^2\theta_{12}=\dd\frac{1}{2}-3\dd\frac{\delta v_T}{v_T}
+\dd\frac{1}{4(\bar{a} b+\bar{b} a+\vert b\vert^2)}
\left[(2 \bar{a}+\bar{b})(-2\dza+\dzb+\dzc)+c.c.\right]
\label{t212c}
\ee
\be
\tan^2\theta_{23}=1+4\dd\frac{\delta v_T}{v_T}
-\dd\frac{2}{3(\bar{a} b+\bar{b} a-\vert b\vert^2)}[\bar{b}(\dzb-\dzc)+c.c.]
-\dd\frac{1}{(\bar{a} b+\bar{b} a)}[\bar{b}(\dze-\dzf)+c.c.]
\label{t223c}
\ee
Given the expected range for the VEVs for $v_T$, $v_S$ and $u$, 
eq. (\ref{range}), we see that all the
corrections can be kept small, below the percent level. We see that deviations from the leading order predictions are obtained for all measurable quantities at approximately the same level. If we require that the subleading
terms do not spoil the leading order picture, these deviations should not be larger than
about 0.05. This can be inferred by the agreement of  both $\Delta m^2_{sol}/\Delta m^2_{atm}$
and $\tan^2\theta_{12}$ with the experimental values. We then go back to eq. (\ref{range}) and 
include this constraint:
\be
0.0022< \frac{v_S}{\Lambda}\approx \frac{v_T}{\Lambda}\approx \frac{u}{\Lambda}<0.05
\label{range2}
\ee
We recall that the lower bound 0.0022 was derived from the requirement that the Yukawa coupling $y_\tau$ is small enough to justify a truncated perturbative expansion. Note that the ratio $\langle \theta \rangle / \Lambda$ is not limited by the constraint in eq. (\ref{range2}).

%
\section{See-saw realization}
We can easily modify the previous model to acomodate the see-saw mechanism.
We introduce conjugate right-handed neutrino fields $\nu^c$ transforming as a triplet of $A_4$
and we modify the transformation law of the other fields according to the following table:
\\[0.2cm]
\begin{center}
\begin{tabular}{|c||c||c|c|c||c|c|}
\hline
{\tt Field}& $\nu^c$ & $\varphi_S$ & $\xi$ & $\tilde{\xi}$ & $\varphi_0^S$ & $\xi_0$\\
\hline
$A_4$ & $3$ & $3$ & $1$ & $1$ & $3$ & $1$\\ 
\hline
$Z_3$ & $\omega^2$ & $\omega^2$ & $\omega^2$ & $\omega^2$ &  $\omega^2$ & $\omega^2$\\
\hline
$U(1)_R$ & $1$ & $0$ & $0$ & $0$ & $2$ & $2$\\
\hline
\end{tabular}
\end{center}
\vspace{0.2cm}
The superpotential becomes
\be
w=w_l+w_d
\ee 
where the `driving' part is unchanged, whereas $w_l$ is now given by:
\be
w_l=y_e e^c (\varphi_T l)+y_\mu \mu^c (\varphi_T l)'+
y_\tau \tau^c (\varphi_T l)''+ y (\nu^c l)+
(x_A\xi+\tilde{x}_A\tilde{\xi}) (\nu^c\nu^c)+x_B (\varphi_S \nu^c\nu^c)+h.c.+...
\label{wlss}
\ee
dots denoting higher-order contributions. The vacuum alignment proceeds exactly as
discussed in section 4 and also the charged lepton sector is unaffected by the modifications.
In the neutrino sector, after electroweak and $A_4$ symmetry breaking we have Dirac
and Majorana masses:
\be
m^D_\nu=y v_u {\bf 1}~~~,~~~~~~~~~~
M=\left(
\begin{array}{ccc}
A+2 B/3& -B/3& -B/3\\
-B/3& 2B/3& A-B/3\\
-B/3& A-B/3& 2 B/3
\end{array}
\right) u ~~~,
\ee
where ${\bf 1}$ is the unit 3$\times$3 matrix and 
\be
A\equiv 2 x_A ~~~,~~~~~~~B\equiv 2 x_B \frac{v_S}{u}~~~.
\label{add}
\ee
The mass matrix for light neutrinos is $m_\nu=(m^D_\nu)^T M^{-1} m^D_\nu$ with eigenvalues
\be
m_1=\frac{y^2}{A+B}\frac{v_u^2}{u}~~~,~~~~~~~
m_2=\frac{y^2}{A}\frac{v_u^2}{u}~~~,~~~~~~~
m_3=\frac{y^2}{-A+B}\frac{v_u^2}{u}~~~.
\ee
The mixing matrix is the HPS one, eq. (\ref{HPSmatrix}).
In the presence of a see-saw mechanism both normal and inverted hierarchies 
in the neutrino mass spectrum can be realized. If we call $\Phi$ the relative phase
between the complex number $A$ and $B$, then $\cos\Phi>-|B|/2|A|$
is required to have $|m_2|>|m_1|$. In the interval
$-|B|/2|A|<\cos\Phi\le 0$, the spectrum is of inverted hierarchy type,
whereas in $|B|/2|A|\le \cos\Phi\le 1$ the neutrino hierachy is of normal type.
The quantity $|B|/2|A|$ cannot be too large, otherwise the ratio $r$
cannot be reproduced. When $|B|\ll|A|$ the spectrum is quasi degenerate.
When $|B|\approx |A|$ we obtain the strongest hierarchy.
For instance, if $B=-2A+z$ ($|z|\ll |A|,|B|$), we find the following
spectrum:
\be
|m_1|^2\approx \Delta m_{atm}^2(\frac{9}{8}+\frac{1}{12} r)~~~,~~~~~~~
|m_2|^2\approx \Delta m_{atm}^2(\frac{9}{8}+\frac{13}{12} r)~~~,~~~~~~~
|m_3|^2\approx \Delta m_{atm}^2(\frac{1}{8}+\frac{1}{12} r)~~~.
\ee
When $B=A+z$ ($|z|\ll |A|,|B|$), we obtain:
\be
|m_1|^2\approx \Delta m_{atm}^2(\frac{1}{3} r)~~~,~~~~~~~
|m_2|^2\approx \Delta m_{atm}^2(\frac{4}{3} r)~~~,~~~~~~~
|m_3|^2\approx \Delta m_{atm}^2(1-\frac{1}{3} r)~~~.
\ee
These results can be affected by higher-order corrections induced
by non renormalizable operators. As before, charged lepton masses
and mixing angles are unaffected at first order. Dirac and Majorana neutrino 
mass terms are instead corrected at first order, through the insertion
of $\varphi_T/\Lambda$ and also the VEVs receive a corrections.
It is interesting to note that the contribution to
the light neutrino masses coming directly from local operators
of the type $(ll h_u h_u...)$
is highly suppressed compared to the see-saw contribution. The latter is of order 1/VEV, whereas the 
former, due to the $Z_3$ assignment, is of order ${\rm VEV}^2/\Lambda^3$. 
In conclusion, the symmetry structure of our model is fully compatible
with the see-saw mechanism.

%
\section{Quarks}
There are several possibilities to include quarks. At first sight the most appealing
one is to adopt for quarks the same classification scheme under $A_4$ that we have 
used for leptons. Thus we tentatively assume that left-handed quark doublets $q$ transform
as a triplet $3$, while the right-handed quarks $(u^c,d^c)$,
$(c^c,s^c)$ and $(t^c,b^c)$ transform as $1$, $1''$ and $1'$, respectively. We can similarly
extend to quarks the transformations of $Z_3$ and U(1)$_R$ given for leptons in the table
of section 4. The superpotential for quarks reads:
\be
{w}_{q}=
y_d d^c (\varphi_T q)+y_s s^c (\varphi_T q)'+y_b b^c (\varphi_T q)''+
y_u u^c (\varphi_T q)+y_c c^c (\varphi_T q)'+y_t t^c (\varphi_T q)''
+h.c.+...
\label{wq}
\ee
It is interesting to note that such an extrapolation to quarks leads to a diagonal CKM 
mixing matrix in first approximation \cite{ma1,ma1.5}. In fact, starting from eq. (\ref{wq}) and proceeding as 
described in detail for the lepton sector, we  see that the up quark and down quark mass matrices 
are separately diagonal with mass eigenvalues which are left unspecified by $A_4$ and with a hierarchy 
that could be accomodated by a suitable U(1)$_F$ set of charge assignments for quarks. Thus 
the $V_{CKM}$ matrix is the identity in leading order, providing a good
first order approximation.

The problems come when we discuss non-leading corrections. As seen in section 5, 
first-order corrections to the lepton sector should be typically below 0.05,
approximately the square of the Cabibbo angle. Therefore it seems difficult to
reproduce the quark mixing between the first two generations in this scheme,
without introducing new ingredients in the symmetry breaking sector.
Also, by inspecting these corrections more closely, we see that, exactly as in the case
of charged leptons, the quark mass matrices are not modified to first order by
higher dimensional Yukawa operators. The only possible first order changes could
only come from the new vacuum, eq. (\ref{alisub}). Unfortunately, up to very small terms
of order $y_{u(d)}^2/y_{t(b)}^2$ and $y_{c(s)}^2/y_{t(b)}^2$, these corrections
are the same in the up and down sectors - see eq. (\ref{Ue}) - and therefore they almost
exactly cancel in the mixing matrix $V_{CKM}$. We conclude that, if one insists in 
adopting for quarks the same flavour properties as for leptons, than
new sources of $A_4$ breaking are needed in order to produce an acceptable $V_{CKM}$.   

An other point of view is to regard $A_4$ as a special feature of the lepton sector,
and to provide an independent description for quarks.
For instance one could take quarks as invariant under $A_4$ and charged only 
with respect to the U(1) part of the flavour group which controls the mass hierarchy.  
Masses and mixing angles for quarks would emerge from the symmetry breaking of
an abelian continuous flavour symmetry, as in many models of fermion masses \cite{review}.
This possibility has the obvious disadvantage of preventing a unified description
of both quarks and lepton masses, as expected for instance in grand unified theories. 

%
\section{Relation with the Modular Group}
There is an interesting relation between the $A_4$ model considered so far and the modular group.
The modular group is the group of linear fractional transformations acting on a complex variable $z$:
\be
z\to\frac{az+b}{cz+d}~~~,~~~~~~~ad-bc=1~~~,
\label{frac}
\ee
where $a,b,c,d$ are integers. These transformations can be represented by the matrices
\be
\left(
\begin{array}{cc}
a&b\cr
c&d
\end{array}
\right)
\ee
with integer coefficients and determinant 1, belonging to the group $SL(2,Z)$.
Since however a matrix and its opposite in $SL(2,Z)$ define the same linear fractional
transformation, the modular group $\Gamma$ coincides with 
$PSL(2,Z)$, the matrices in $SL(2,Z)$ up to an overall sign.
There are infinite elements in $\Gamma$, but all of them can be generated by the two
transformations:
\be
s:~~~z\to -\frac{1}{z}~~~,~~~~~~~t:~~~z\to z+1~~~,
\label{st}
\ee
represented by the matrices  \footnote{Notice that $m_t$ is not unitary and also that $m_s^2=-1$ while $m_t^n\ne1$ for all integer $n>1$.}:
\be
m_s=
\left(
\begin{array}{cc}
0&1\cr
-1&0
\end{array}
\right)~~~,~~~~~~~~~~
m_t=
\left(
\begin{array}{cc}
1&1\cr
0&1
\end{array}
\right)~~~.
\ee
In string theory the transformations (\ref{st})
operate in many different contexts. For instance the role of the complex 
variable $z$ can be played by a field, whose VEV can be related to a physical
quantity like a compactification radius or a coupling constant. In that case
$s$ in eq. (\ref{st}) represents a duality transformation and $t$ in eq. (\ref{st}) represent the transformation 
associated to an ''axionic'' symmetry. 

The transformations $s$ and $t$ in (\ref{st}) satisfy the relations
\be
s^2=(st)^3=1
\label{absdef}
\ee
and, conversely, these relations provide an abstract characterization of the modular group.
Since the relations (\ref{$A_4$}) are a particular case of the more general constraint (\ref{absdef}),
it is clear that the representations given in (\ref{s$A_4$}) and (\ref{ST}) are also representations of the modular
group. Then the natural questions is: how much special are (\ref{s$A_4$}) and (\ref{ST}) from the point of view
of $\Gamma$? To anwser this we should inspect the unitary irreducible representations of $\Gamma$.
To this purpose it is sufficient to look at linear unitary irreducible representations of the two elements
$s$ and $t$, from which it is possible to reconstruct the representatives of any other element in $\Gamma$.

The singlet representations act simply as multiplications by a phase factor:
\be
s=e^{\dd i 2\pi\alpha}~~~,~~~~~~~~~~t=e^{\dd i 2 \pi\beta}
\label{albe}
\ee
It is immediate to see that there are six inequivalent choices for $\alpha$ and $\beta$ that satisfy
(\ref{absdef}):
\be
\begin{array}{lll}
1&s=1&t=1\\
1^I&s=1&t=e^{\dd i 4 \pi/3}\\
1^{II}&s=1&t=e^{\dd i 2\pi/3}\\
1^{III}&s=-1&t=e^{\dd i 5\pi/3}\\
1^{IV}&s=-1&t=e^{\dd i 3\pi/3}\\
1^V&s=-1&t=e^{\dd i \pi/3}
\label{singl}
\end{array}
\ee
The representation 1 is the invariant representation. If we have two fields $a$ and $b$ transforming according
two representations of the above list, then their product $a b$ also transforms according to a singlet representation
and the multiplication table can be easily deduced by eqs. (\ref{singl}):
\be
\begin{array}{l|llllll}
&1&1^I&1^{II}&1^{III}&1^{IV}&1^{V}\\
\hline
1&1&1^I&1^{II}&1^{III}&1^{IV}&1^{V}\\
1^I&1^I&1^{II}&1&1^{IV}&1^V&1^{III}\\
1^{II}&1^{II}&1&1^I&1^V&1^{III}&1^{IV}\\
1^{III}&1^{III}&1^{IV}&1^{V}&1^{I}&1^{II}&1\\
1^{IV}&1^{IV}&1^V&1^{III}&1^{II}&1&1^I\\
1^{V}&1^V&1^{III}&1^{IV}&1&1^I&1^{II}
\end{array}
\ee
In particular we see that, beyond the invariant
representation, there is an interesting subset of representations closed under the product. It is
given by $1$, $1^I$, $1^{II}$. These representations corresponds to the representations
$1$, $1'$ and $1''$ used in our model.

\begin{figure}[h!]
\centerline{\psfig{file=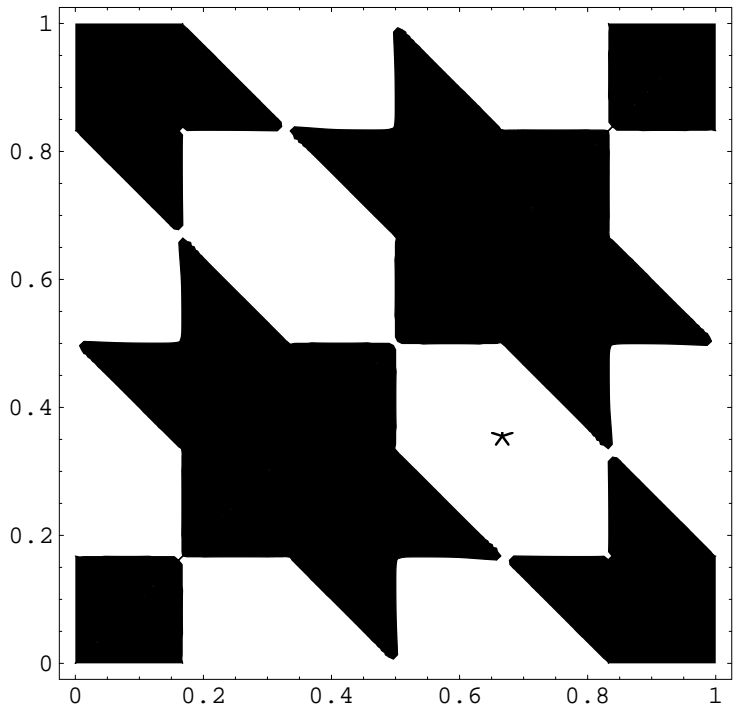,width=0.6\textwidth}}
\caption{Allowed (white) region in the parameter space $(\beta_2,\beta_3)$. The conditions 
(\ref{con1},\ref{con2}) of appendix A
are saturated along the lines $\beta_2=1/6,3/6,5/6$, $\beta_3=1/6,3/6,5/6$, $\beta_3=-\beta_2+3/6$,
$\beta_3=-\beta_2+5/6$, $\beta_3=-\beta_2+7/6$, $\beta_3=-\beta_2+9/6$. Accounting for the periodicity
in $\beta_{2,3}$, we have six allowed regions with the same shape and area, differing only
by the location of their center. These six regions are related by permutations of $\beta_1$, $\beta_2$
and $\beta_3$ and the inequivalent representations can be labelled by the points
of one of them. The irreducible representation used in our model is marked by a star.}
\end{figure}

The relations $s^2=(st)^3=1$ lead to a quantization of $\alpha$ and $\beta$ in eq. (\ref{albe}) and,
in the unidimensional case,  there is a finite number of unitary representations. 
This is no-longer true when going to higher-dimensional representations \cite{tuba}.
We are particularly interested in the three-dimensional case which is discussed in detail in
the appendix A. The result is that we have unitary inequivalent representations described by
\be
t=
\left(
\begin{array}{ccc}
e^{\dd i 2 \pi \beta_1}&0&0\\
0&e^{\dd i 2 \pi \beta_2}&0\\
0&0&e^{\dd i 2 \pi \beta_3}
\end{array}
\right)~~~,
\ee
and a matrix $s$ also determined as function of $\beta_i$ once the following relations are staisfied: $\beta_1+\beta_2+\beta_3=0$ (mod 1) and $(\beta_2,\beta_3)$ in the region
$3/6\le\beta_2\le 5/6$, $1/6\le\beta_3\le 3/6$, $-\beta_2+5/6\le\beta_3\le-\beta_2+7/6$.
Therefore, there is a double infinity of possible three-dimensional representations 
which in the $(\beta_2,\beta_3)$ plane appears as in fig. 1.
The representation (\ref{ST}) adopted in the construction of the model falls just in the center of the 
region allowed to $(\beta_2,\beta_3)$. Perhaps in the underlying theory there is a dynamical principle that selects this particularly symmetric point. 

\section{Conclusion}

The $A_4$ discrete group appears to be particularly suitable to economically reproduce the precise relations among neutrino mass matrix elements which are needed to obtain the HPS mixing matrix. We have presented here a 4-dimensional SUSY version of an $A_4$ model which is completely natural, in the sense that no arbitrary tuning of parameters is necessary to lead to the HPS mixing angles, once the ratios of VEVs to the cutoff are fixed in a given interval. A moderate fine tuning is only present in the neutrino mass spectrum, in order to reproduce the observed small value of $r=\Delta m_{sun}^2/\Delta m_{atm}^2 $. In a previous paper we had presented a model which was equally natural, actually in a more extended interval for the cut-off, but assumed extra dimensions, thus making the model more exotic. We then pointed out an interesting way of presenting the $A_4$ group as a particular set of transformations of the modular group. This approach has the immediate advantage that the lagrangian basis where the symmetry is formulated coincides with the basis where the charged leptons are diagonal.  But this connection could possibly lead to an insight on the possible origin of the $A_4$ symmetry within the context of a more fundamental theory. Finally, if the same structure of left-handed and right-handed field classification in $A_4$ is extended from leptons to quarks, then, in leading order, the CKM matrix coincides with the unit matrix. A study of non leading corrections in this model shows that the departures from unity of the CKM matrix are far too small to reproduce the observed mixing angles. Thus the quark mixing angles, in this picture, should arise from additional effects specific of the quark sector.

\section*{Acknowledgements}

We thank  S. Ferrara, W. Lerche, Y. Lin and I. Masina for very interesting discussions. We are particularly indebted with G.G. Ross for pointing out to us the importance of an additional flavon singlet for obtaining the correct vacuum alignment. 

\vfill
\newpage

\section*{Appendix A}
We explicitly classify and build the three dimensional, irreducible
unitary representations of the modular group.
It is not restrictive to go to a basis where $T$ is diagonal:
\be
T=
\left(
\begin{array}{ccc}
t_1&0&0\\
0&t_2&0\\
0&0&t_3
\end{array}
\right)\equiv
\left(
\begin{array}{ccc}
e^{\dd i 2 \pi \beta_1}&0&0\\
0&e^{\dd i 2 \pi \beta_2}&0\\
0&0&e^{\dd i 2 \pi \beta_3}
\end{array}
\right)~~~.
\ee
In this basis and without loss of generality, the most general 3 by 3 unitary matrix 
$S$ satisfying $S^2=1$ can be parametrized as:
\be
S=\eta
\left(
\begin{array}{ccc}
-\Cc&\Sc\Sp&\Sc\Cp\\
\Sc\Sp&\Cc \SSp-\CCp&(1+\Cc)\Cp \Sp\\
\Sc \Cp&(1+\Cc)\Cp \Sp&\Cc \CCp-\SSp
\end{array}
\right)~~~,
\ee
where $\eta=\pm 1$. We can take $\chi$
between 0 and $\pi$, and $\varphi$ between 0 and $\pi/2$
\footnote{It is not restrictive to assume the non-diagonal elements of $S$
real and of the same sign. If this is not the case, we can make them
real and of the same sign by means of a unitary diagonal transformation
that leaves $T$ invariant and does not affect the diagonal elements of $S$.}.
We first discuss $\eta=1$.
The condition $(ST)^3=1$, written in the form $T S T = S T^{-1}S$, 
gives rise to the equations:
\be
\begin{array}{l}
-\dd\frac{\CCc}{t_1}-\Cc~ t_1^2-\dd\frac{\SSc}{t_2 t_3}(\CCp~ t_2+\SSp~t_3)=0\\
-\dd\frac{\SSc\SSp}{t_1}-\dd\frac{(\CCp-\Cc\SSp)^2}{t_2}
+(\Cc \SSp-\CCp) t_2^2\\
-\dd\frac{(1+\Cc)^2\CCp\SSp}{t_3}=0\\
-\dd\frac{\SSc\CCp}{t_1}-\dd\frac{(\SSp-\Cc\CCp)^2}{t_3}
+(\Cc \CCp-\SSp) t_3^2\\
-\dd\frac{(1+\Cc)^2\CCp\SSp}{t_2}=0\\
t_1t_2+\CCp (\dd\frac{1}{t_2}-\dd\frac{1}{t_3})+\Cc(\dd\frac{1}{t_1}
-\dd\frac{\SSp}{t_2}-\dd\frac{\CCp}{t_3})=0\\
t_1 t_3 +\SSp(-\dd\frac{1}{t_2}+\dd\frac{1}{t_3})+\Cc(\dd\frac{1}{t_1}
-\dd\frac{\SSp}{t_2}-\dd\frac{\CCp}{t_3})=0\\
-\dd\frac{\SSc}{t_1}+\dd\frac{(1+\Cc)}{t_2 t_3}(\SSp~ t_2+\CCp~ t_3+t_2^2 t_3^2-\Cc(\CCp~ t_2+\SSp~ t_3))=0
\end{array}
\label{dim3}
\ee
By combining the fourth and the fifth equations above we obtain:
\be
t_1 t_2 t_3=1~~~,
\label{ttt}
\ee
which can be solved, for instance, to express $t_1$ in terms of $t_2$ and $t_3$.
By ignoring solutions that lead to reducible representations, the remaining independent 
conditions in (\ref{dim3}) are solved by:
\be
\begin{array}{c}
\Cc=-\dd\frac{t_2 t_3 (t_2+t_3)}{1-t_2^2 t_3-t_2 t_3^2+t_2^3 t_3^3}\\
\\
\Cp=\sqrt{\dd\frac{t_2(1+t_3^3)(1-t_2^2 t_3)}{(t_2-t_3)(1+t_2^3t_3^3-2t_2 t_3(t_2+t_3))}}~~~.
\label{chiephi}
\end{array}
\ee
It is easy to see that $\Cc$ is automatically real and that the condition $\vert \Cc \vert\le 1$
reads:
\be
t_2^3 t_3^3+\dd\frac{1}{t_2^3 t_3^3}-2 t_2 t_3(t_2+t_3)-\dd\frac{2}{t_2 t_3}
(\dd\frac{1}{t_2}+\dd\frac{1}{t_3})+2\ge 0~~~.
\label{con1}
\ee
The argument of the squared root in eq. (\ref{chiephi}) is always real and 
the conditions to obtain $\Cp$ real and bounded between 0 and 1 are, respectively:
\be
\begin{array}{c}
-\dd\frac{2(1+t_3^3)}{t_2^3 t_3^4} (t_2-t_3+2 t_2 t_3^3+t_2^2 t_3^2-3 t_2^3 t_3
-3 t_2^3 t_3^4+t_2^4 t_3^3+2 t_2^5 t_3^2+t_2^5 t_3^5-t_2^6 t_3^4)\ge 0\\
-\dd\frac{2(1+t_2^3)}{t_2^4 t_3^3} (t_3-t_2+2 t_3 t_2^3+t_3^2 t_2^2-3 t_3^3 t_2
-3 t_3^3 t_2^4+t_3^4 t_2^3+2 t_3^5 t_2^2+t_3^5 t_2^5-t_3^6 t_2^4)\ge 0
\end{array}
\label{con2}
\ee
Recalling that $t_{2,3}=e^{\dd i 2\pi\beta_{2,3}}$, we find that in the fundamental region
$0\le \beta_{2,3}\le 1$ the values of $\beta_{2,3}$ that are compatible with the conditions
(\ref{con1},\ref{con2}) are those displayed in figure 1. Then $t_1$ is given by the constraint 
(\ref{ttt}).
The representation chosen in our model, corresponds to one of the centers of the allowed regions, 
the point $(\beta_2,\beta_3)=(2/3,1/3)$. Then eq. (\ref{chiephi}) gives
$\Cc=1/3$, $\Cp=1/\sqrt{2}$ and $\Sc=2\sqrt{2}/3$, $\Sp=1/\sqrt{2}$.

Finally, the irreducible representations corresponding to $\eta=-1$, which flips the sign of $S$, are obtained
from those given above by sending $(S,T)$ into $(-S,-T)$.

\vfill
\newpage

\section*{Appendix B}
Here we discuss how the vacuum alignment achieved at the leading order is modified
by the inclusion of higher dimensionality operators.
The part of the superpotential depending on the driving fields $\varphi_0^T$, $\varphi_0^S$ and $\xi_0$
is modified into
\be
w_d+\Delta w_d
\ee
Here $w_d$ is the leading order contribution:
\bea
w_d&=&M (\varphi_0^T \varphi_T)+ g (\varphi_0^T \varphi_T\varphi_T)\nn\\
&+&g_1 (\varphi_0^S \varphi_S\varphi_S)+
g_2 \tilde{\xi} (\varphi_0^S \varphi_S)+
3 \tilde{g}_3^2 \xi_0 (\varphi_S\varphi_S)-
\tilde{g}^2_4 \xi_0 \xi^2+
g_5 \xi_0 \xi \tilde{\xi}+
g_6 \xi_0 \tilde{\xi}^2~~~,
\label{wdApp}
\eea
where, for convenience, we have redefined $g_3\equiv 3\tilde{g}_3^2$ and $g_4\equiv -g_4^2$. We recall that $w_d$ gives rise
to the munimum:
\be
\begin{array}{lr}
\varphi_T=(v_T,0,0)&v_T=-\dd\frac{3M}{2g}\\
\varphi_S=(v_S,v_S,v_S)&v_S=\dd\frac{\tilde{g}_4}{3 \tilde{g}_3} u\\
\xi=u&\\
\tilde{\xi}=0&
\label{sols}
\end{array}
\ee
with $u$ undetermined. The remaining term, $\Delta w_d$ is the most general quartic, $A_4$-invariant polynomial
linear in the driving fields: 
\be
\Delta w_d=\dd\frac{1}{\Lambda}\left(
\sum_{k=3}^{13} t_k I_k^T+
\sum_{k=1}^{12} s_k I_k^S+
\sum_{k=1}^{3} x_k I_k^X
\right)
\ee
where $t_k$, $s_k$ and $x_k$ are coefficients and $\{I_k^T,I_k^S,I_k^X\}$ represent a basis of independent quartic invariants: 
\be
\begin{array}{ll}
I_3^T=(\varphi_0^T\varphi_T) (\varphi_T\varphi_T)&
I_9^T=\left(\varphi_0^T(\varphi_S\varphi_S)_S\right) \xi\\
I_4^T=(\varphi_0^T\varphi_T)' (\varphi_T\varphi_T)''&
I_{10}^T=\left(\varphi_0^T(\varphi_S\varphi_S)_S\right) \tilde{\xi}\\
I_5^T=(\varphi_0^T\varphi_T)'' (\varphi_T\varphi_T)'&
I_{11}^T=(\varphi_0^T\varphi_S) \xi^2\\
I_6^T=(\varphi_0^T\varphi_S) (\varphi_S\varphi_S)&
I_{12}^T=(\varphi_0^T\varphi_S) \xi \tilde{\xi}\\
I_7^T=(\varphi_0^T\varphi_S)' (\varphi_S\varphi_S)''&
I_{13}^T=(\varphi_0^T\varphi_S) {\tilde{\xi}}^2\\
I_8^T=(\varphi_0^T\varphi_S)'' (\varphi_S\varphi_S)'
\end{array}
\ee
\be
\begin{array}{ll}
I_1^S=\left((\varphi_0^S\varphi_T)_S(\varphi_S\varphi_S)_S\right)&
I_7^S=\left(\varphi_0^S(\varphi_T\varphi_S)_S\right) \tilde{\xi}\\
I_2^S=\left((\varphi_0^S\varphi_T)_A(\varphi_S\varphi_S)_S\right)&
I_8^S=\left(\varphi_0^S(\varphi_T\varphi_S)_A\right) \xi\\
I_3^S=(\varphi_0^S\varphi_T) (\varphi_S\varphi_S)&
I_9^S=\left(\varphi_0^S(\varphi_T\varphi_S)_A\right) \tilde{\xi}\\
I_4^S=(\varphi_0^S\varphi_T)' (\varphi_S\varphi_S)''&
I_{10}^S=(\varphi_0^S\varphi_T) \xi^2\\
I_5^S=(\varphi_0^S\varphi_T)'' (\varphi_S\varphi_S)'&
I_{11}^S=(\varphi_0^S\varphi_T) \xi \tilde{\xi}\\
I_6^S=\left(\varphi_0^S(\varphi_T\varphi_S)_S\right) \xi&
I_{12}^S=(\varphi_0^S\varphi_T) {\tilde{\xi}}^2
\end{array}
\ee
\be
\begin{array}{ll}
I_1^X=\xi_0 \left(\varphi_T(\varphi_S\varphi_S)_S\right)&
I_3^X=\xi_0 (\varphi_S\varphi_S) \tilde{\xi}\\
I_2^X=\xi_0 (\varphi_S\varphi_S) \xi
\end{array}
\ee
The new minimum for $\varphi_S$, $\varphi_T$, $\xi$ and $\tilde{\xi}$ is obtained by searching
for the zeros of the $F$ terms, the first derivatives of $w_d+\Delta w_d$, associated to 
the driving fields $\varphi_0^S$, $\varphi_0^T$ and $\xi_0$.
We look for a solution that perturbes (\ref{sols}) to first order in the $1/\Lambda$
expansion:
\be
\begin{array}{lr}
\varphi_T=(v_T+\delta v^T_1,\delta v^T_2,\delta v^T_3)&v_T=-\dd\frac{3M}{2g}\\
\varphi_S=(v_S+\delta v_1,v_S+\delta v_2,v_S+\delta v_3)&v_S=\dd\frac{\tilde{g}_4}{3 \tilde{g}_3} u\\
\xi=u&\\
\tilde{\xi}=\delta u'
\label{sols1}
\end{array}
\ee
The minimum conditions become equations in the unknown $\delta v^T_i$, $\delta v_i$, $u$ and $\delta u'$
which can be expanded in $1/\Lambda$. By keeping only the first order in the expansion, we get:
\be
\begin{array}{l}
-\delta v^T_1-\dd\frac{27 t_3}{8 g^3}\dd\frac{M^2}{\Lambda}+
\dd\frac{\tilde{g_4}}{3 \tilde{g_3}} 
\left(t_{11}+\dd\frac{\tilde{g_4}^2}{3 \tilde{g_3}^2}(t_6+t_7+t_8)\right)
\dd\frac{u^3}{M \Lambda}=0\\
2 \delta v^T_3+
\dd\frac{\tilde{g_4}}{3 \tilde{g_3}} 
\left(t_{11}+\dd\frac{\tilde{g_4}^2}{3 \tilde{g_3}^2}(t_6+t_7+t_8)\right)
\dd\frac{u^3}{M \Lambda}=0\\
2 \delta v^T_2+
\dd\frac{\tilde{g_4}}{3 \tilde{g_3}} 
\left(t_{11}+\dd\frac{\tilde{g_4}^2}{3 \tilde{g_3}^2}(t_6+t_7+t_8)\right)
\dd\frac{u^3}{M \Lambda}=0\\
\dd\frac{g_2 \tilde{g_4}}{3 \tilde{g_3}} \delta u'
+\dd\frac{2 g_1 \tilde{g_4}}{9 \tilde{g_3}} 
\left(2 \delta v_1-\delta v_2-\delta v_3
\right)
-\dd\frac{1}{g}\left(\dd\frac{3}{2}s_{10}+\frac{\tilde{g_4}^2}{2 \tilde{g_3}^2}s_3+\dd\frac{\tilde{g_4}}{3 \tilde{g_3}} s_6\right)
\dd\frac{M u}{\Lambda}=0\\
\dd\frac{g_2 \tilde{g_4}}{3 \tilde{g_3}} \delta u'
+\dd\frac{2 g_1 \tilde{g_4}}{9 \tilde{g_3}} 
\left(2 \delta v_2-\delta v_1-\delta v_3
\right)
-\dd\frac{1}{g}\left(\frac{\tilde{g_4}^2}{2 \tilde{g_3}^2}s_4-\dd\frac{\tilde{g_4}}{\tilde{g_3}} (\dd\frac{s_6}{6}+\dd\frac{s_8}{4})\right)
\dd\frac{M u}{\Lambda}=0\\
\dd\frac{g_2 \tilde{g_4}}{3 \tilde{g_3}} \delta u'
+\dd\frac{2 g_1 \tilde{g_4}}{9 \tilde{g_3}} 
\left(2 \delta v_3-\delta v_1-\delta v_2
\right)
-\dd\frac{1}{g}\left(\frac{\tilde{g_4}^2}{2 \tilde{g_3}^2}s_5-\dd\frac{\tilde{g_4}}{\tilde{g_3}} (\dd\frac{s_6}{6}-\dd\frac{s_8}{4})\right)
\dd\frac{M u}{\Lambda}=0\\
g_5 \delta u'+2 \tilde{g_3}\tilde{g_4}(\delta v_1+\delta v_2+\delta v_3)-\dd\frac{\tilde{g_4}}{2 g \tilde{g_3}} x_2 
\dd\frac{M u}{\Lambda}=0
\end{array}
\ee
These equations are solved by:
\be
\begin{array}{l}
\delta v^T_1=-\dd\frac{3 t_3}{2 g}\dd\frac{v_T^2}{\Lambda}+\left[
-\dd\frac{\tilde{g_4}}{2 g \tilde{g_3} t_{11}}
-\dd\frac{\tilde{g_4}^3}{6 g \tilde{g_3}^3}(t_6+t_7+t_8)\right] \dd\frac{u^3}{v_T\Lambda}\\
\delta v^T_2=\delta v^T_3=\left[
\dd\frac{\tilde{g_4}}{4 g \tilde{g_3} t_{11}}
+\dd\frac{\tilde{g_4}^3}{12 g \tilde{g_3}^3}(t_6+t_7+t_8)\right] \dd\frac{u^3}{v_T\Lambda}\\
\delta v_1=\left[
\dd\frac{g_5 s_{10}}{6 g_2\tilde{g_4}^2}
+\dd\frac{g_5}{18 g_2\tilde{g_3}^2}(s_3+s_4+s_5)
-\dd\frac{\tilde{g_3 s_{10}}}{g_1 \tilde{g_4}}
-\dd\frac{\tilde{g_4}}{6 g_1 \tilde{g_3}}(2 s_3-s_4-s_5)
-\dd\frac{s_6}{3 g_1}
-\dd\frac{x_2}{18 \tilde{g_3}^2}
\right]\dd\frac{v_T u}{\Lambda}\\
\delta v_2=\left[
\dd\frac{g_5 s_{10}}{6 g_2\tilde{g_4}^2}
+\dd\frac{g_5}{18 g_2\tilde{g_3}^2}(s_3+s_4+s_5)
+\dd\frac{\tilde{g_3 s_{10}}}{2 g_1 \tilde{g_4}}
-\dd\frac{\tilde{g_4}}{6 g_1 \tilde{g_3}}(2 s_4-s_3-s_5)
+\dd\frac{s_6}{6 g_1}+\dd\frac{s_8}{4 g_1}
-\dd\frac{x_2}{18 \tilde{g_3}^2}
\right]\dd\frac{v_T u}{\Lambda}\\
\delta v_3=\left[
\dd\frac{g_5 s_{10}}{6 g_2\tilde{g_4}^2}
+\dd\frac{g_5}{18 g_2\tilde{g_3}^2}(s_3+s_4+s_5)
+\dd\frac{\tilde{g_3 s_{10}}}{2 g_1 \tilde{g_4}}
-\dd\frac{\tilde{g_4}}{6 g_1 \tilde{g_3}}(2 s_5-s_3-s_4)
+\dd\frac{s_6}{6 g_1}-\dd\frac{s_8}{4 g_1}
-\dd\frac{x_2}{18 \tilde{g_3}^2}
\right]\dd\frac{v_T u}{\Lambda}\\
\delta u'=-\left[
\dd\frac{\tilde{g_3 s_{10}}}{g_2 \tilde{g_4}}
+\dd\frac{\tilde{g_4}}{3 g_2 \tilde{g_3}}(s_3+s_4+s_5)
\right]\dd\frac{v_T u}{\Lambda}~~~,
\end{array}
\ee
where $u$ remains undetermined.
This justifies the corrections (\ref{alisub}) to the vacuum alignment adopted in section 5.

\end{document}